# Network Lifetime Analysis of AODV, DSR and ZRP at Different Network Parameters

Niranjan Kumar Ray[1], Harsh Bardhan Sharma [2] and Ashok Kumar Turuk [3]

Department of Computer Science and Engineering, National Institute of Technology, Rourkela, India
`rayniranjan@gmail.com.`[1]`,hs8906@gmail.com,`[2] `akturuk@nitrkl.ac.in`[3]

## ABSTRACT

*Enhancement of network lifetime is a key design criterion for most of the energy constrained networks as nodes are battery operated. In multi-hop wireless network, proper utilization of battery power is very much necessary to maintain network connectivity. If the battery power of a node drains quickly then its connectivity in its neighborhood will be lost. So the study of network lifetime is very much crucial as compared to other network parameters. Considering this importance we made an attempt to study the behaviour of three most common routing protocols in ad hoc network. Extensive simulations are done on AODV, DSR and ZRP to determine the network lifetime at different node mobility and at different network load. Simulation results suggest that AODV is the most energy efficient protocol as compared to other.*

## 1. INTRODUCTION

Wireless technology has strongly influenced our personal and professional lives in the recent time due to it's applicability and versatility in different fields. It enhanced our computing, communication skills and information accessing capabilities through many modern types of equipment. Numerous wireless technologies in form of cellular phones, PDA, blue-tooth devices, and many hand-held computers, are extending the wireless communication to a fully pervasive computing environment as a result this is emerging as one of the most pervasive computing technologies. Wireless network can be classified to infrastructure supported network or infrastructure independent networks. The former requires specific network backbone in form of access points or base stations to support communication, while the latter doesn't need such for it's operation and popularly known as wireless ad hoc network. Multi hop wireless networks in all their different forms such as mobile ad hoc network (MANET) and vehicular ad hoc network (VANET), wireless sensor network(WSN), wireless mesh network(WMN), etc are coming under this category. In multi-hop ad hoc network destination nodes may be multiple hops away from the source node. This approach provides a number of advantages as compare to single-hop networking solution. Some of its advantages are (i) support for self configuration and adaption at low cost, (iii) support of load balancing for increasing network life, (iv) greater network flexibility, connectivity, etc[1], [2], [3]. However irrespective of these advantages it also suffered with many challenges associated with restricted battery capacity, unpredictable mobility, routing, etc. [4]. The basic requirement in MANET is how to deliver packets efficiently among the mobile nodes. Since node's topology changes frequently this makes routing very problematic. Also low bandwidth, limited battery capacity and error prone medium adds further complexity in designing an efficient routing protocol. Estimation of end-to-end delays, efficient utilization of bandwidth, and proper use of available resource are the most common requirements as resource allocation; capacity planning, file sharing, etc depend on them. Also power managements is very much necessary as nodes are battery operated and it is always a cumbersome task to recharge or replace the battery as the network is deployed in such environments where it is neither feasible nor economical to perform that task. Bad carrier sensing, retransmission due to collision of packets, exchange of large number of control message to find paths are some of the major cause of energy consumption. With a proper analysis of battery consumptions, light weight applications, efficient network protocol and interface power consumption of wireless network can be properly addressed. Flooding based routing protocols rely on message forwarding by broadcasting the message. This mechanism consumes a major portion of battery power at node level also affect the longevity of the network. Energy efficient routing protocols apply some techniques to reduce flooding mechanism by some probabilistic and heuristic based approach but are suffered with increase end-to-end delay and decrease network throughput. For this reason there must be some threshold between power consumptions and other network parameters while designing routing protocols for MANET. In the literature different techniques are proposed to find the energy efficiency of routing protocol, but network lifetime is not properly addressed at different network traffic, load and mobility. Focusing on these three parameters we made an attempt to determine the network lifetime of AODV, DSR and ZRP at variable speed and load. AODV and DSR represent the reactive category of routing mechanism and ZRP represent the hybrid approach of routings routing in ad hoc network.

The rest of the paper is organised as follows. In Section 2, we present routing protocols of MANETs under which we discuss about AODV, DSR and ZRP. In Section 3, we discuss the network lifetime parameter and simulation environments. In Section 4 simulation results are discussed at different network conditions. In Section 5, we end our discussion with conclusion and thought for future work on this topic.

## 2. ROUTING PROTOCOLS

Routing protocols in MANET based on their functionalities are classified to (i) Reactive (ii) Proactive and (iii) Hybrid. Reactive protocols established path based on the present requirements for which they known as on-demand routing protocol. Proactive protocols in other hand obtain the path by the help of routing table information. Routing tables are periodically updated. Hybrid protocols carry some feature from both categories. Reactive protocols are considered the most suitable for network with higher mobility as compare to proactive protocols. Proactive protocols are best fit to the static network where node information does not change frequently.

We have considered the basic routing protocol of MANET for our study e.g. Ad hoc on demand distance vector (AODV), Destination sequence routing (DSR) and Zone routing protocol (ZRP). AODV and DSR are reactive routing protocols while ZRP belongs to hybrid routing protocol.

Perkin et al. first presented AODV in [5] which then standardize in IETF RFC 3561 in 2003[6]. It is an improvement on the DSDV [7] algorithm and modified message broadcasting procedure to minimize the number of required broadcasts by creating routes on a demand basis instead of maintaining a complete list of routes as observed in the DSDV algorithm. AODV makes use of destination sequence numbers to ensure all routes are loop-free and contain the most recent route information. The path discovery process is initiated when a source node wishes to send a message to a destination node and does not have a valid route to destination. The source broadcasts a route request (RREQ), recipients of RREQ perform the same work if they don't know the path to destination. The forwarding continues till message reach the destination. If intermediate node finds a fresh enough route in its route cache then rather than forwarding it reply back to the source. Each node maintains its own sequence number and broadcast ID. The broadcast ID is incremented for every RREQ the node initiates. RREQ are uniquely identified with the sequence number and source node's address. The initiator along with its sequence number and broadcast ID includes the most recent sequence number it has for the destination in the RREQ. Intermediate nodes reply to the RREQ with a RREP packet having similar fields as that of RREQ packet, only if they have a route to the destination whose corresponding destination sequence number is equal or greater than that contained in the RREQ. When forwarding the RREQ, intermediate nodes records in their routing tables the address of the neighbors from which the first copy of the broadcast packet is received, thereby establishing a reverse path. If additional copies of the same RREQ are later received, these packets are discarded. A timer is associated with each route cache entry which will cause the deletion of the entry if it is not used within the specified time as a result, AODV supports only symmetric links. As the RREP travels back to the source using the reverse path, all intermediate nodes create a new route entry for destination.

In DSR, when node needs to send a packet to another node it places a complete list of hops to follow in its actual data packet header, if it has the route to the destination. It first checks its route cache and if doesn't find the destination in its route cache it start the proceedings to find the path. If it has a route to the destination, it will use this route to send the packets. Otherwise node initiates the route discovery by broadcasting a RREQ packet. RREQ contains sender and destination address. Intermediate node on receiving the packet checks destination address, if it knows the destination address it reply to the source. Otherwise it appends its address to the route record and broadcasts it further. A node discards the request it has recently seen another route request from the same initiator or the route record already contains its own address. In this way it prevents looping. The Route maintenance is accomplished through the use of route error packets and acknowledgements. Each node broadcast packets and waits for acknowledgements. If a node does not receive an acknowledgement after forwarding a packet, it sends an acknowledgement request for certain times, if it doesn't received any reply, the node generates a route-error (RERR) message and sends it to the sender of packet. ZRP [9] is the first hybrid category protocol for ad-hoc mobile network. It effectively combines best features reactive and proactive protocols. It employed the concept of proactive routing scheme within limited zone within the r-hop neighborhood of each node, and use reactive approach beyond that zone. It uses two routing schemes: Intra-zone routing protocol (IARP) and an inter-zone routing protocol (IERP). IARP is used to maintain routing information and provides route to nodes within zone. IERP uses the route query (RREQ)/ route reply (RREP) packets to discover a route in a way similar to typical on-demand routing protocols. In ZRP, a routing zone consists of a few nodes within one, two, or a couple of hops away from each other. It works similar to a clustering with the exception that every node acts as a cluster head and a member of other clusters. Each zone has a predefined zone centred at itself in terms of number of hops. Within this zone a table-driven-based routing protocol is used. This implies that route updates are performed for nodes within the zone. Therefore, each node has a route to all other nodes within its zone. If the destination node resides outside the source zone, an on-demand search-query routing method is used. It also uses Border cast Resolution Protocol (BRP). When intended destination is not known, RREQ packet is broadcast via the nodes on the border of the zone. Route queries are only broadcast from one node's border nodes to other border nodes until one node knows the exact path to the destination node. ZRP limits the proactive overhead to only the size of the zone, and the reactive search overhead to only selected border nodes. The IARP in ZRP must be able to determine a node's neighbors itself. This protocol is usually a pro-active protocol and is responsible for the routes to the peripheral nodes.

## 3. PERFORMANCE EVALUATION

To study the performance of MANET routing protocols researchers have focused basically on throughput, end to end delay and packet delivery ratio, jitters, hop count metrics etc. [11][12][15][16]. However network life time is not properly addressed nor presented with greater accuracy. Considering the importance in this paper we measure the network lifetime of AODV, DSR and ZRP at different node density, traffic connection and in different mobility pattern.

### 3.1. Simulator

Qualnet network simulator [10] is used here for simulating multihop ad hoc wireless network routing protocols. It is a commercial version of GloMoSim [17] and is developed by Scalable Network Technology. It is extremely scalable and can supports high fidelity models of networks of thousands of nodes. In a discrete event simulator like Qualnet, the simulation performed only when an event occurs. It is based on event scheduler, which contains any events that needs to be processed and stepped through. Processing of an event may produce some new events. These new events are inserted into event scheduler. QualNet is modelling software that predicts performance of networking protocols and networks through simulation and emulation. The simulator is written in C++ while it's graphical toolkits are implemented in Java. Every protocols in QualNet starts with an initialization function which takes an inputs and configures the protocol. Event dispatcher activated when an event occurs in a layer. The simulator checks the type of event and calls appropriate event handler to process. At the end of the simulation finalization function is called to print out the collected statistics. Figure 1 shows the event handling process in QualNet.

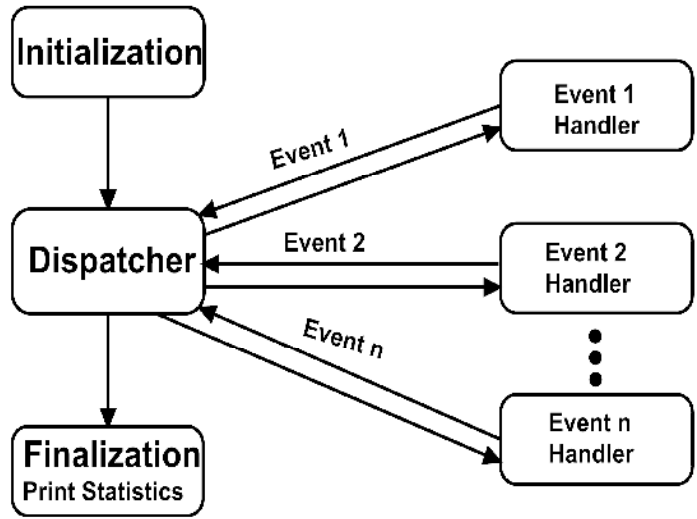

Figure 1 Event handling process

QualNet Developer is another tool used for modelling large wired and wireless networks. It helps to design network scenarios, developing new protocols, optimizing new and existing models etc. The network architecture of developer is shown in figure 2. It consists of scenario designer to design new scenarios in graphical mode, analyzer tools for statistical analysis of simulation results, command line interface for compiling and debugging [18].

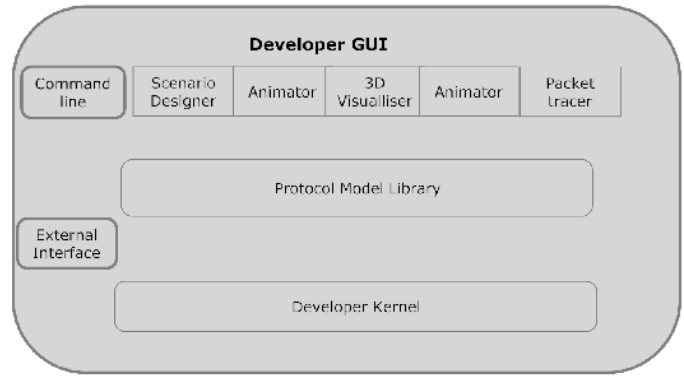

Figure 2 Developer architecture

## 3.2 Network Environment

We consider a network of square area of (1500m × 1500m) where nodes deployed uniformly. The simulation parameters are shown in table 1.

| Table 1: Simulation Parameters | |
|---|---|
| Network Simulator | QualNet 4.5.1 |
| Size of the physical terrain | Dimension ( 1500, 1500 ) mts |
| Coordinate System | Cartesian |
| Node Positioning | Uniform |
| Channel frequency | 2.4 GHz |
| Propagation path loss model | Two-Ray |
| Energy Model | User defined |
| Battery model | Simple linear |
| MAC-protocol | IEEE 802.11b |
| Routing policy | Dynamic |

We consider energy model whose energy power amplification efficiency factor is 6.5. Transmit circuit power consumptions is taken as 220 mw and receive power consumption is 200 mw. Idle power consumption is considered as 150 mw and supply voltage is 3V. IEEE 802.11 DCF is used as the MAC protocol. The radio model uses bit rate of 2 Mbps and has a radio range of 250 meters. We generated 90 different scenarios with varying movement patterns and communication patterns. Since each method was changed in an identical fashion for which direct comparisons are made from simulator. For the above comparison of routing protocols constant bit rate (CBR) traffic patterns are used. The network contains variable CBR traffic connections and packet size of 512 bytes. Packets are send from source nodes in the 1s interval.

International Journal of Mobile Network Communications & Telematics (IJMNCT) Vol.2, No.3, June 2012

### 3.3. Performance Metrics

We considered the network lifetime of the node at different network scenarios. The network life time is defined as:

It is the time until first battery drain out which is same as that of minimum life time of overall nodes in the network.

$$LT_{Network} = \min_{u \in N}(LT_u)$$

Where $LT_{Network}$ and $LT_u$ represents the network life time and node life time respectively. The node life time under a given flow can be represented as:

$$LT_u = \frac{E_u}{\sum_{v \in n_u^1} e_{uv} \sum q_{uv}}$$

Where

$E_u$ = Initial battery power of node u

$e_{uv}$ = transmission energy required to reach from node u to v

$q_{uv}$ = The rate at which packet transmitted from u to v

$n_u^1$ = One hop neighbor list of node u.

## 4. RESULT DISCUSSIONS

For the measurement of network lifetime we generate different scenarios varying the number of nodes, CBR connection and pause time. Analyses are done below in respective sections.

### 4.1. Network Lifetime at Varying Mobility Model

We simulate the network at varying node of 30, 40 and 50. Extensive simulations are done to find the network lifetime of AODV, DSR and ZRP at three mobility model. We consider Random way point mobility (RWP), Group mobility and NONE (no mobility). In order to find the best mobility model we fix the CBR connection and pause time of each node. We found through simulation as well as from the theoretical study that RWP performed better as compared to other two as gives higher network lifetime at different node density. Also we found that irrespective of type of mobility pattern AODV gives better network life in all scenarios. Figure 3, 4 and 5 shows the lifetime analysis of AODV, DSR and ZRP at node 30, 40 and 50 respectively. In the rest of our analysis we only consider RWP mobility due to its superior results.

42

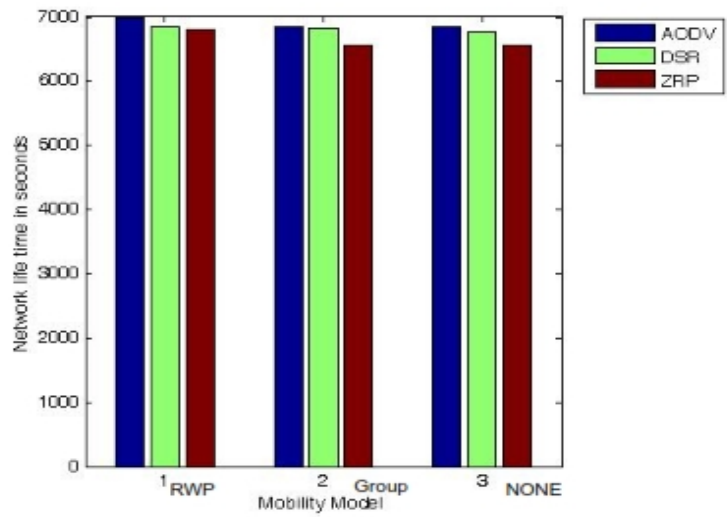

Figure 3. Network lifetime with 30 nodes

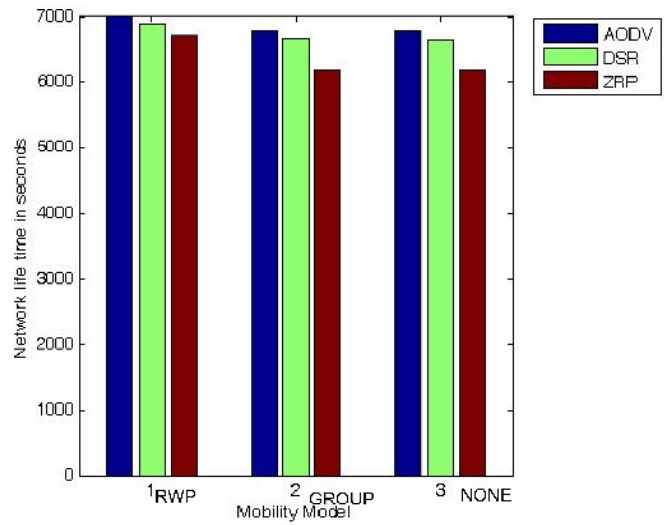

Figure 4. Network lifetime with 40 nodes

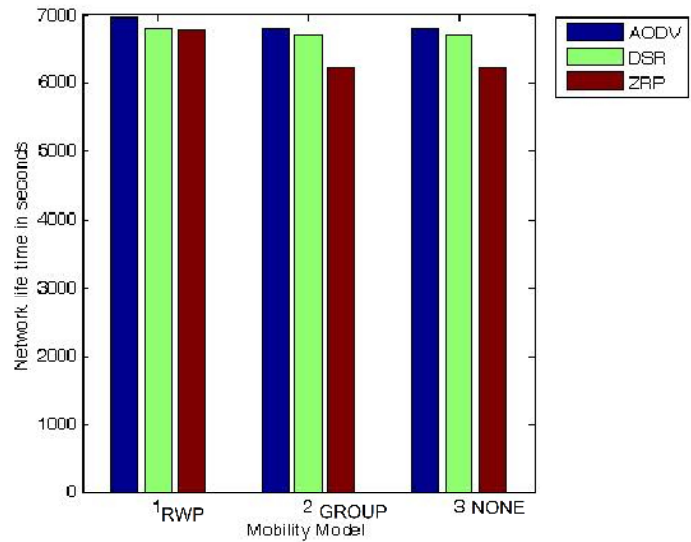

Figure 5. Network lifetime with 50 Nodes

## 4.2. Network Lifetime at Varying CBR Connections

The Figure 6 to 8 shows network lifetime at 30, 40 and 50 nodes with varying CBR connections. For this purpose we start the simulations with 30 nodes with initial six source destination connections. Then we gradually increase the connections by adding two more connections each time up to 14 different source-destination connections. We study the behaviour between network traffic and load. In all case AODV is performs better. The life time of ZRP is always low as compare to DSR and AODV. The improve performance of AODV is its greater compatibility with mobility network size.

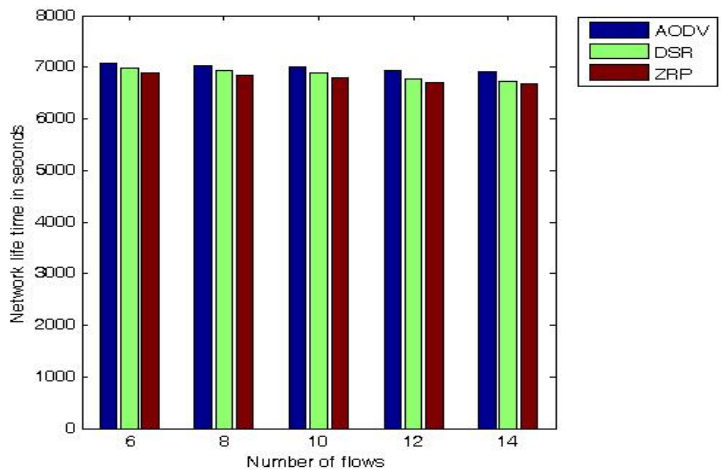

Figure 6. Network lifetime with 30 nodes

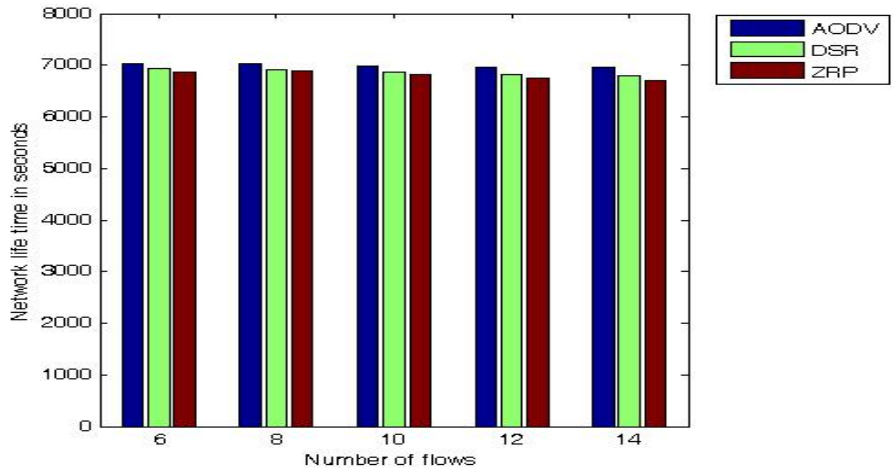

Figure 7. Network lifetime with 40 nodes

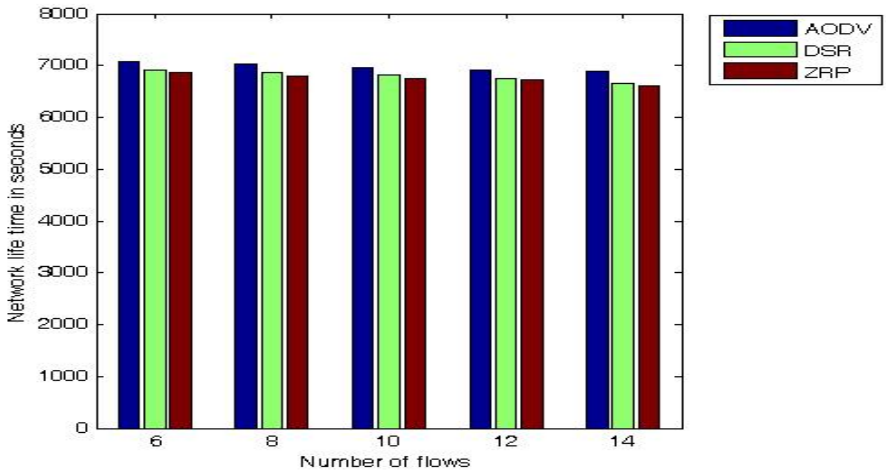

Figure 8. Network lifetime with 50 nodes

### 4.3. Network Lifetime at Varying Pause Time

The mobility plays a vital role in the network life time. We have already discusses the simulation results of mobility models. For pause time variations we consider RWP for our comparisons. In RWP model node remains stationary for a certain periods of time (known as pause time). At the end of that time node choose for a random destination in (1500m × 1500m) simulation space area. The node moves to the destination at a speed in the range [0, max]. When node reaches the destination it waits for time equal to pause time and starts moving for another destination. It repeats this behaviour for the entire simulation time. We simulate with four different pause times: 0s, 10s, 20s, and 30s. The pause time of 0 seconds represent the continuous movements. As the movement pattern is very sensitive to performance measurement, we generate 36 different scenarios with simulator. Figure 9 to 11 shows the network life time at various pause times at different node density. In all results it is also found that AODV performing better as compared to

45

International Journal of Mobile Network Communications & Telematics (IJMNCT) Vol.2, No.3, June 2012

DSR and ZRP. It is observed that pause time of the node doesn't make more impact on the life time. It indicates irrespective of change in pause time all the protocols battery drain time not changing more. In this simulation pause time is varied by keeping maximum node speed as a constant. Figure 9 to 11 shows the lifetime of three routing protocol at different pause times.

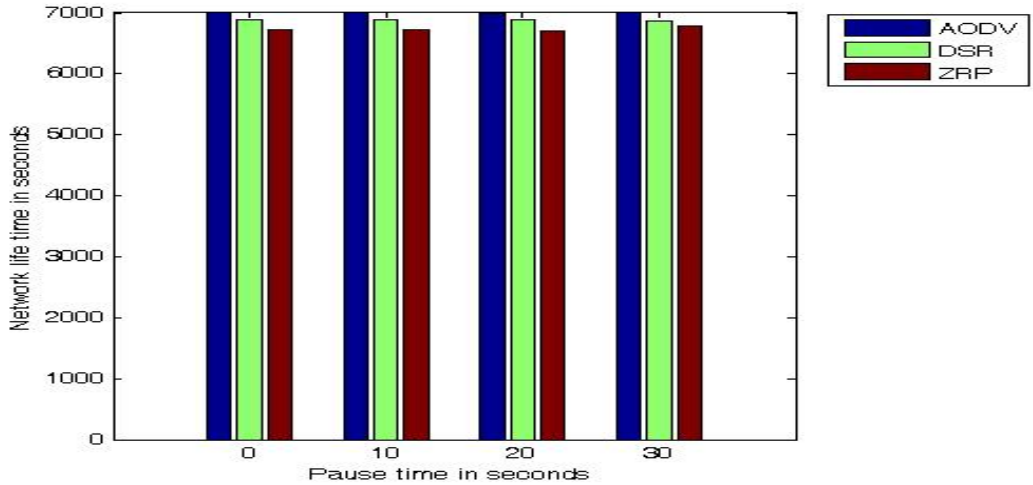

Figure 9. Network lifetime with 30 nodes

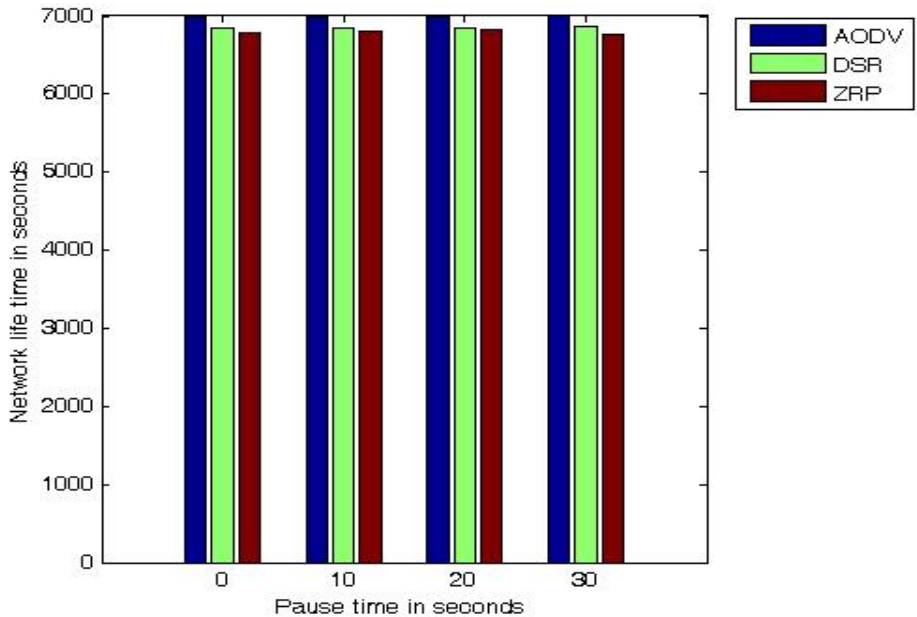

Figure 10. Network lifetime with 40 nodes

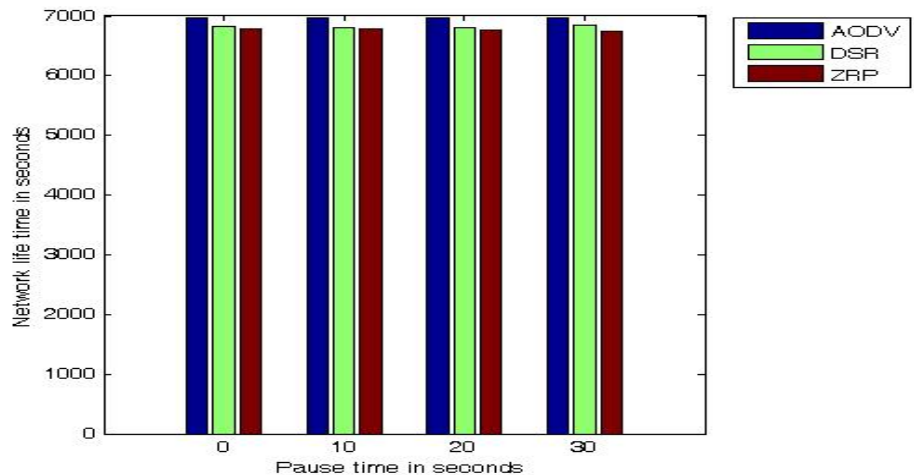

Figure 11. Network lifetime with 50 nodes

## 5. CONCLUSIONS

This paper compares the network lifetime of three most common routing protocols in wireless ad hoc networks. Network lifetime always play a vital role in determining the network connectivity and network capacity of energy constraints network [13, 14]. In order to provide a complete analysis for determining network lifetime efficient protocol we consider node mobility, node density and network traffic parameters in this study. It observed that lifetime of the ADOV is better as compared to than DSR and ZRP. In these three network condition it is found that AODV is always superior in term of node mobility, increase traffic and increase node density. With this network lifetime analysis we agreed with other study that AODV is the standard routing protocol for energy constraint MANETs. In this work we have also considered the network life time of a node with different mobility model (e.g. group mobility, None mobility, RWP) but we have not measure the lifetime at variable connections (e.g. ftp, telnet). Our future work will focus on these parameters to analyze network life time. We are presently working on theoretical analysis of network lifetime.

## ACKNOWLEDGEMENTS

Authors would like to thank the anonymous reviewers for their constructive comments on the manuscripts. This work was supported by Department of Information Technology, Govt. of India under the grant no: 13(1)/2008-CC&BT.

## Authors

**Niranjan Kumar Ray** received his MTech degree in computer scienc from Utkal University, Bhubaneswar in 2007. He has been a PhD student in the department of Computer Science and Engineering at National Institute of Technology, Rourkela since 2008. His research interest includes wireless ad hoc and sensor networks, at present he is working on MAC and routing protocols for energy effciency. He is a student member of IEEE.

**Harsh Bardhan Sharma** is a MTech student in the Department of Computer Science and Engineering at National Institute of Technology, Rourkela since 2010.

**Ashok Kumar Turuk** received his PhD degree in Computer Science and Engineering from IIT Kharagpur in 2005. He joined National Institute of Technology, Rourkela (Formerly REC Rourkela) as a faculty in the department of Computer Science and Engineering in 1997 and presently held the position of Associates Professor and working as the head of the department. He published more than 60 different papers in journals and conference proceedings. His research interest includes wireles ad hoc, sensor, and optical networks. He was the program chair of of International Conference on Communication Computing and Security (ICCCS-2011). He is the principal investigator (PI) of different projects funded by Government agencys like DST and DIT. He is a member of IEEE Computer Scociety, Indian Society for Technical Education (ISTE), Computer Society of India (CSI) and Orissa Bigyan Academy.